\documentclass{IEEEtran4PSCC}

\usepackage{cite}

\newcommand{\myquad}[1][1]{\hspace*{#1em}\ignorespaces}

\ifCLASSINFOpdf
   \usepackage[pdftex]{graphicx}
\else
   \usepackage[dvips]{graphicx}
\fi
\usepackage[inkscapelatex=false]{svg}

\usepackage[cmex10]{amsmath}
\usepackage{bbm}
\usepackage{siunitx}

\usepackage{array}
\makeatletter
\let\old@ps@headings\ps@headings
\let\old@ps@IEEEtitlepagestyle\ps@IEEEtitlepagestyle
\def\psccfooter#1{%
    \def\ps@headings{%
        \old@ps@headings%
        \def\@oddfoot{\strut\hfill#1\hfill\strut}%
        \def\@evenfoot{\strut\hfill#1\hfill\strut}%
    }%
    \def\ps@IEEEtitlepagestyle{%
        \old@ps@IEEEtitlepagestyle%
        \def\@oddfoot{\strut\hfill#1\hfill\strut}%
        \def\@evenfoot{\strut\hfill#1\hfill\strut}%
    }%
    \ps@headings%
}
\makeatother

\psccfooter{%
        \parbox{\textwidth}{\hrulefill \\ \small{24th Power Systems Computation Conference} \hfill \begin{minipage}{0.2\textwidth}\centering \vspace*{4pt} \includegraphics[scale=0.06]{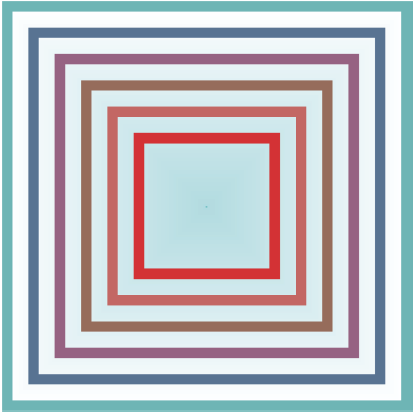}\\\small{PSCC 2026} \end{minipage} \hfill \small{Limassol, Cyprus --- June 8-12, 2026}}%
}

\begin{document}
%
\title{Data-driven Communication and Control Design for Distributed Frequency Regulation with Black-box Inverters}


\author{\IEEEauthorblockN{Michael C. A. Nestor\IEEEauthorrefmark{1},
Jiaxin Wang\IEEEauthorrefmark{2},
Ning Zhang\IEEEauthorrefmark{2}}
and Fei Teng\IEEEauthorrefmark{1}

\IEEEauthorblockA{\IEEEauthorrefmark{1} Department of Electrical and Electronic Engineering,
Imperial College London,
London, U.K.}
\IEEEauthorblockA{\IEEEauthorrefmark{2} Department of Electrical Engineering,
Tsinghua University,
Beijing, China\\
m.nestor22@ic.ac.uk, jiaxinwangthu@gmail.com, ningzhang@tsinghua.edu.cn, f.teng@ic.ac.uk}
}

\maketitle

\begin{abstract}
The increasing penetration of inverter-based resources into the power grid, with often only black-box models available, challenges long-standing frequency control methods. Most recent works take a decentralized approach without online device coordination via communication. This paper considers both dynamic behavior and communication within secondary frequency control on an intermediate timescale. We develop a distributed data-driven approach that utilizes peer-to-peer communication between inverters to avoid the need for a central control center. To enable a trade off between communication network requirements and control performance, we present a framework to guide communication topology design for secondary frequency regulation. Following design of the inter-agent information exchange scheme, we design a controller that is structured according to the communication topology with a closed-loop stability guarantee. Case studies on the IEEE 39-bus system validate the framework and illustrate the trade-off between communication requirements and control performance that is enabled by our approach.

\end{abstract}


\thanksto{\noindent Submitted to the 24th Power Systems Computation Conference (PSCC 2026).}

\section{Introduction}
\label{section:introduction}

Power systems are undergoing a global transformation from fossil fuel-based to renewable generation. Renewable energy sources (RES) such as wind and solar photovoltaic (PV) are typically interfaced to the grid using inverters, as are storage devices such as batteries and a growing proportion of loads. Inverter behavior is largely defined by software rather than inherent physical dynamics, enabling an entirely new approach to controlling and managing power grids. Their fast dynamic capability allows for more rapid control updates to adjust to changing grid conditions and reduced system inertia whilst the increasing number of controllable devices points towards a more scalable approach to grid control compared to the traditional centralized methodology. Meanwhile, manufacturers may not share proprietary inverter models with system operators making only black-box models available, meaning that traditional model-based grid control may not be applicable; data-driven control methods compatible with black-box models have gained increasing traction in recent years \cite{Markovsky-2023-DeePC-Tutorial}.\par

In this paper, we address secondary frequency regulation within the black-box inverter paradigm through a distributed data-driven control strategy. We aim to design a stabilizing optimal regulator that will steer the bus frequencies back to their nominal values following a disturbance. This can be achieved by collecting persistently exciting data \cite{Willems-2005-Fundamental-Lemma} from the power system and designing the controller directly from the dataset without an intermediate identification step. The controller is implemented in a distributed fashion; each power system bus computes its secondary regulation input locally, using local information and information received via communication from other control agents. This avoids the need for a central control center to handle all communications with participating devices. In order to reduce the level of communication necessary, we design an inter-agent information exchange scheme to determine the communication neighborhoods of each agent, balancing communication overheads against response optimality whilst ensuring closed-loop stability is achievable. Compared to traditional secondary regulation, our approach can operate on a faster timescale (order of 1 \si{\second}) and can be executed simultaneously with primary control of synchronous generators (SGs) to take advantage of the fast control capabilities of inverter-based resources (IBRs), thus requiring system operators to account for the secondary control layer when assessing system security.

\subsection{Related Work}

Traditional approaches to secondary frequency regulation in transmission systems, known as Automatic Generation Control (AGC), use an integral control approach to regulate the power system to zero deviation in steady-state frequency and tie-line power flows based on an Area Control Error (ACE) for each power system area \cite{Kumar-Kothari-2005-AGC-Strategies}. The standard architecture for AGC has been relatively unchanged since the 1950s and does not leverage subsequent advances in communication and computational capabilities \cite{Dorfler-2019-Distributed-Control-Optimization-Autonomous-Grids}. Due to the growing penetration of distributed generation, the topic of distributed secondary frequency regulation utilizing peer-to-peer communication has attracted considerable attention \cite{Molzahn-2017-Survey-Distributed-Optimization-Control-Algorithms} \cite{Wu-2016-Enhanced-Secondary-Frequency-Control-Distributed-P2P-Communication}. Model-based strategies proposed in the literature include distributed model predictive control (MPC) \cite{Venkat-2008-Distributed-MPC-AGC}, primal-dual gradient descent methods \cite{Wang-Chen-2025-Distributed-Coordination-GFM-GFL-Optimal-Frequency-Control}, a virtual swing equation with decentralized dissipativity conditions \cite{Kasis-2019-Stability-Optimality-Distributed-Secondary-Frequency-Control} and distributed-averaging proportional-integral control \cite{Simpson-Porco-2021-On-Stability-of-DAPI-Frequency-Control}. For such methods relying on system models, it has been noted that inaccurate nominal models can degrade dynamic performance and even cause instability \cite{Kumar-Kothari-2005-AGC-Strategies}; alongside the lack of white-box inverter models this has led to interest in model-free control approaches. Proposed schemes include feedback optimization \cite{Zholbaryssov--Dominguez-Garcia-2021-Safe-Data-Driven-Secondary-Control-DERs}, which may require sufficient timescale separation between physical and optimization dynamics, and data-driven disturbance estimation \cite{Ekomwenrenren-2024-Data-Driven-FFC-IBRs} with upper-layer distributed coordination between system areas \cite{Ekomwenrenren-2021-Hierarchical-Coordinated-FFC-Using-IBRs}; dynamic interactions between areas do not appear to be directly considered. The communication scheme utilized in a distributed control approach will impact performance; for example, due to latency. The co-optimization of communication resource allocation and frequency reserves can improve operational profitability \cite{Zhang-2025-Pricing-Framework-Activation-Latency-FFR-VPP}, whilst communication latencies have been shown to impact the performance and stability of AGC schemes based on 5G wireless communication \cite{He-2024-Impact-Comms-Delay-5G-Freq-Regulation-Performance}. Previous work has developed a model-based framework for designing the sparsity of the communication network by optimizing the controller structure for wide-area frequency control \cite{Dorfler-et-al-2014-Sparsity-Promoting-Optimal-WAC}. Investigating the design of the inter-agent communication topology within distributed data-driven secondary frequency control therefore appears to be a timely avenue of research.

\subsection{Main Contributions}

In this paper, we formulate secondary frequency regulation as a linear control design problem. We develop a fully data-driven scheme for designing a communication topology and a distributed controller structured according to the resulting communication topology that implements secondary control in transmission systems. Our approach guarantees frequency stability in closed-loop and zero steady-state frequency error assuming linearized dynamics. No prior knowledge of the grid topology or system parameters such as aggregated bus inertia is necessary for these guarantees to hold, though the approach is able to incorporate prior knowledge if it is available. Furthermore, we utilize a stability constraint in the topology design stage such that we guarantee the feasibility of the subsequent control design problem, assuming that the control designer uses an approach compatible with the topology designer's stability constraint. The separation of these two stages is modular and avoids a monolithic approach to communication and control design. We allow for a trade-off between communication requirements and control performance, thus reducing the cost of communication resources at the expense of a reduction in control performance.

\subsection{Paper Structure}

The power system modelling and communication topology design framework are given in Section \ref{section:problem_formulation}, before our control design and communication topology design schemes are presented in Sections \ref{section:control_design} and \ref{section:topology_design}, respectively. Case studies validating our approach are analyzed in Section \ref{section:simulations} and the paper is concluded in Section \ref{section:conclusion}.

\section{Problem Formulation}
\label{section:problem_formulation}

\subsection{Frequency Dynamics Modelling}



We consider that inertia and damping may be provided either by synchronous generators (SGs) or inverter-based resources (IBRs). This includes grid-forming (GFM) or grid-supporting (e.g., grid-following (GFL) with droop; see \cite{Zhang-2021-GFM-Converters-in-RES-Grid-Strategy-Stability-Applications} and \cite{Ochoa-2025-Optimal-Control-Robust-Dynamic-Performance-Inverter-Dominated-Power-Systems} for a detailed description) inverters, which may be connected to renewable energy sources, storage units or controllable loads. Consider an electric power transmission grid modelled as a graph \(\mathcal{G}_P := \{\mathcal{V},\mathcal{E}_P\}\), where \(\mathcal{V}\) represents the set of buses and \(\mathcal{E}_P\) represents the lines connecting buses. Let \(N_I\) be the number of buses with inertia (and damping) and \(N_L\) be the number without; the total number of buses is \(N_I + N_L\). Denote the set of buses with inertia by \(\mathcal{N} := \{1,\ldots,N_I\}\) and the set without inertia by \(\mathcal{L} := \{N_I+1,\ldots,N_I+N_L\}\) such that \(\mathcal{V} = \mathcal{N} \cup \mathcal{L}\). We can partition the system susceptance matrix accordingly:
\begin{equation}
    B=\begin{bmatrix}
        B_{\mathcal{NN}}&B_{\mathcal{NL}}\\
        B_{\mathcal{LN}}&B_{\mathcal{LL}}
    \end{bmatrix}.
\end{equation}
We assume that the nonlinear power system dynamics are linearized about an operating point, which is an equilibrium of the system, and use \(\Delta\) to denote quantities relative to this equilibrium. The power injection at bus $i\in\mathcal{N}$, \(\Delta p_{i}^{\mathrm{inj}}\), is given by
\begin{equation}
    \Delta p_{i}^{\mathrm{inj}}=\sum_{i=1}^{N_I} J_{ij} \Delta \theta_j + \sum_{\ell=1}^{N_L}L_{i\ell} \Delta p_{N_I+\ell}^{L},
\end{equation}
where \(\Delta \theta_i\) is the power angle deviation at bus \(i\) and
\begin{equation}
    J:=B_{\mathcal{NN}}-B_{\mathcal{NL}}B_{\mathcal{LL}}^{-1}B_{\mathcal{LN}},\ L:=B_{\mathcal{NL}}B_{\mathcal{LL}}^{-1}.
\end{equation}
Neglecting fast electromagnetic and electromechanical dynamics, the swing equation for bus~$i\in\mathcal{N}$ is given by 
\begin{subequations}
    \begin{align}
        \frac{\mathrm{d} \Delta \theta_i}{\mathrm{d}t} &= \omega_0\Delta \omega_i \\
        \begin{split}
            m_i \frac{\mathrm{d} \Delta \omega_i}{\mathrm{d}t} + d_i \Delta \omega_i &= \Delta p_i^{\mathrm{pri}}+\Delta p_i^\mathrm{sec} - \sum_{j=1}^{N_I} J_{ij}\Delta\theta_j \\ 
            - \sum_{\ell=1}^{N_L}  L_{i\ell} \Delta p_{N_I+\ell}^{\mathrm{L}} &- \Delta p_i^{\mathrm{dist}} -\sum_{\ell=1}^{N_L}L_{i\ell}\Delta p_{N_I+\ell}^{\mathrm{dist}},
        \end{split}
    \end{align}
\end{subequations}
where for bus \(i\), \(\Delta \omega_i\) is the bus angular frequency deviation, \(\Delta p_i^\mathrm{pri}\) is the primary control power injection, \(\Delta p_i^\mathrm{sec}\) is the secondary control power injection, $\Delta p_i^{\mathrm{dist}}$ is the disturbance power, \(\Delta p_{i}^{\mathrm{L}}\) is the load damping response, \(m_i\) and \(d_i\) are the bus inertia and damping, respectively, and \(\omega_0\) is the nominal angular frequency. All quantities are expressed per unit and we assume buses with no inertia do not participate in secondary control or have a primary control response.\par
\begin{figure}[h]
    \centering
    \includegraphics[width=0.8\linewidth]{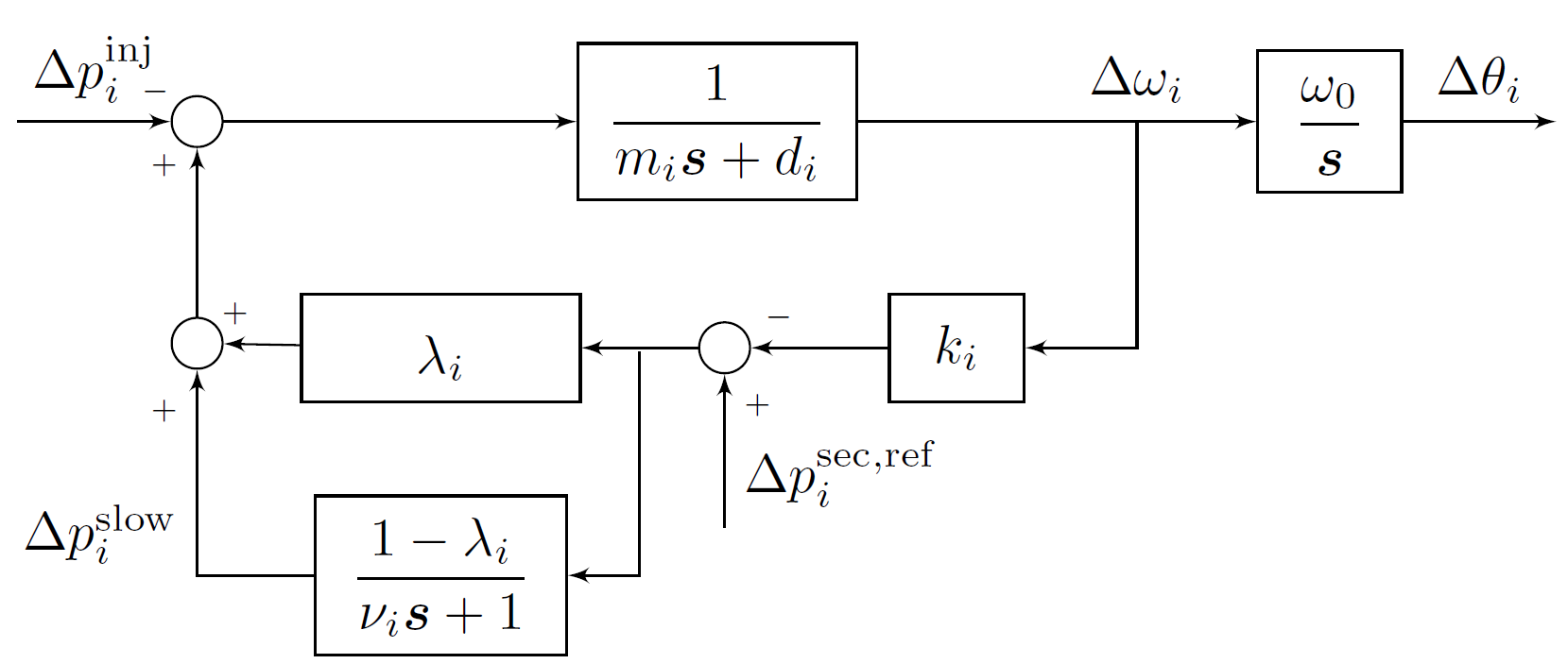}
    \caption{Block diagram of our SG model, showing the primary droop, secondary input, re-heater process and inertial response.}
    \label{fig:SG_block_diagram}
\end{figure}
\begin{figure}[thbp]
    \centering
    \includegraphics[width=0.9\linewidth]{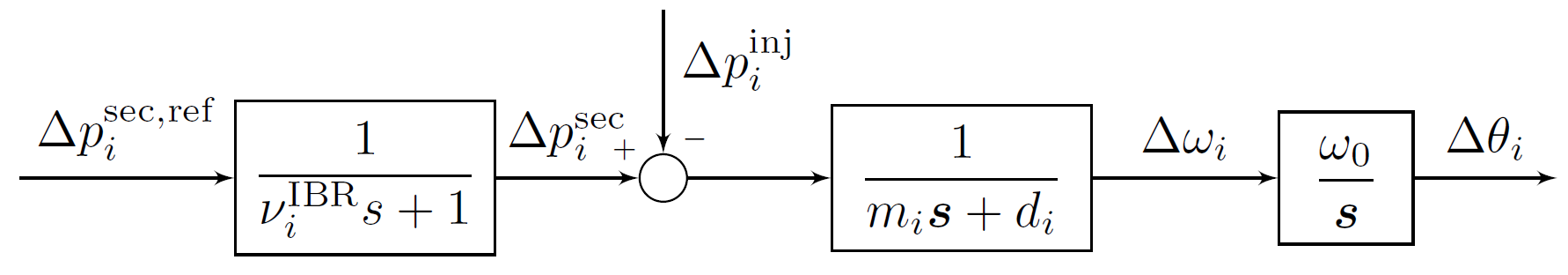}
    \caption{Block diagram of our IBR model with VSG control, showing the secondary input and inertial response.}
    \label{fig:VSG_block_diagram}
\end{figure}
\begin{figure}[h]
    \centering
    \includegraphics[width=0.95\linewidth]{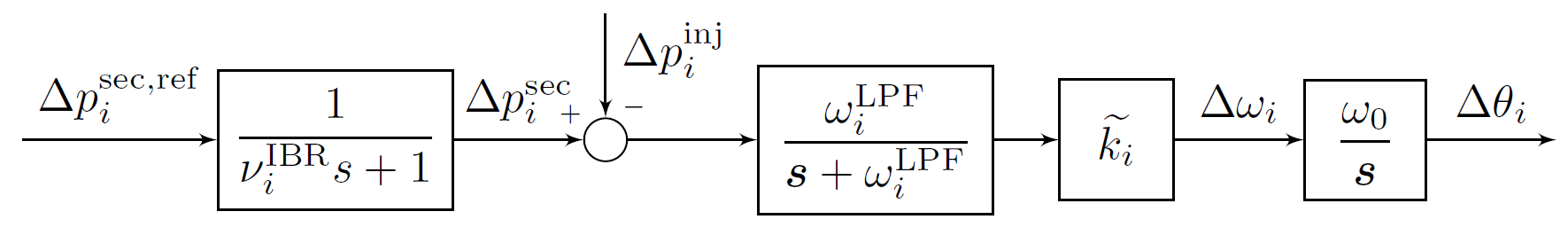}
    \caption{Block diagram of our IBR model with droop control, showing the secondary input response, low-pass filter and droop gain.}
    \label{fig:GFM_droop_block_diagram}
\end{figure}

We model the synchronous generator re-heater process dynamics by:
\begin{equation}
    \Delta p_i^{\mathrm{pri}}(s)=-k_i\frac{\lambda_i\nu_i s +1}{\nu_i s +1} \Delta\omega_i(s),
\end{equation}
where \(k_i\) is the governor droop coefficient, \(\nu_i\) is the slow reheat process time constant and \(\lambda_i \in (0,1)\) is the ratio of the fast reheat loop to the slow reheat loop in terms of mass flow rate. We neglect the turbine and governor dynamics as these are typically faster than the re-heater process. The block diagram representing the SG dynamics is shown in Figure \ref{fig:SG_block_diagram}.
For inverters with droop control and virtual SG (VSG) control, most literature assume that \(\Delta p_i^{\mathrm{pri}}=0\) because they mainly provide virtual inertia and damping before secondary frequency regulation. We assume that a VSG controller calculates the IBR frequency using parameters \(m_i\) and \(d_i\) applied to a first-order inertial response. The block diagram for an IBR with VSG control is shown in Figure \ref{fig:VSG_block_diagram}.
We assume that a GFM inverter with droop control passes the measured difference between the power setpoint and the bus power injection through a low-pass filter with cut-off frequency \(\omega_i^\mathrm{LPF}\) before multiplying by the IBR droop gain \(\Tilde{k}_i\) to calculate the bus frequency. This dynamic response can be expressed using an equivalent inertia and damping, which may be calculated from droop parameters by: \(m_i := \frac{1}{\Tilde{k}_i \omega_i^\mathrm{LPF}}\) and \(d_i := \frac{1}{\Tilde{k}_i}\), respectively. The block diagram for an IBR with droop control is shown in Figure \ref{fig:GFM_droop_block_diagram}.

We model the IBR tracking of the desired secondary input as a first-order response: \(\frac{\mathrm{d} \Delta p_i^\mathrm{sec}}{\mathrm{d}t} = \frac{1}{\nu_i^\mathrm{IBR}} (\Delta p_i^\mathrm{sec,ref} - \Delta p_i^\mathrm{sec})\), where \(\nu_i^\mathrm{IBR}\) is the time constant of the \(i\)-th IBR, \(\Delta p_i^\mathrm{sec,ref}\) is the setpoint provided by the secondary controller and \(\Delta p_i^\mathrm{sec}\) is the actual power output of the IBR.\par

The system dynamics may be formulated as a state-space model:
\begin{subequations}
    \begin{align}
        \Dot{x} = A_cx + B_c u + B_{c,d} \Delta p^\mathrm{dist},
    \end{align}
\end{subequations}
where \(x = \begin{bmatrix}
    \Delta \theta^\top & \Delta \omega^\top & \Delta p^\mathrm{sec^\top} & \Delta p^\mathrm{slow^\top}
\end{bmatrix}^\top \in \mathbbm{R}^n\). We denote \(\Delta \theta = \begin{bmatrix}
    \Delta \theta_i
\end{bmatrix}_{i \in \mathcal{N}}\), \(\Delta \omega = \begin{bmatrix}
    \Delta \omega_i
\end{bmatrix}_{i \in \mathcal{N}}\), \(\Delta p^\mathrm{sec} = \begin{bmatrix}
    \Delta p^\mathrm{sec}_i
\end{bmatrix}_{i \in \mathcal{N}_{\mathrm{IBR}}}\) where \(\mathcal{N}_{\mathrm{IBR}}\) is the set of buses connected to an IBR, \(\Delta p^\mathrm{slow} = \begin{bmatrix}
    \Delta p^\mathrm{slow}_i
\end{bmatrix}_{i \in \mathcal{N}_{\mathrm{SG}}}\) where \(\mathcal{N}_{\mathrm{SG}}\) is the set of buses connected to an SG, \(u = \begin{bmatrix}
    \Delta p^\mathrm{sec,ref}_i
\end{bmatrix}_{i \in \mathcal{N}}\) and \(\Delta p^\mathrm{dist} = \begin{bmatrix}
    \Delta p^\mathrm{dist}_i
\end{bmatrix}_{i \in \mathcal{N}} + L \begin{bmatrix}
    \Delta p^\mathrm{dist}_i
\end{bmatrix}_{i \in \mathcal{L}}\). In particular, the state-space matrices are given by:
\begin{subequations}
    \begin{align}
        A_c &= \begin{bmatrix}
            0 & \omega_0 I_{N_I} & 0 & 0 \\
            -M^{-1} J & -M^{-1} (\Tilde{D} + S) & M_\mathrm{IBR}^{-1} & M_\mathrm{SG}^{-1} \\
            0 & 0 & T_\mathrm{IBR}^{-1} & 0 \\
            0 & -T_\mathrm{SG}^{-1} (I - \Lambda) \Tilde{K}^\mathrm{pri} & 0 & -T_\mathrm{SG}^{-1}
        \end{bmatrix} \\
        B_c &= \begin{bmatrix}
            0 \\ M^{-1} \Theta_\mathrm{SG} \Lambda \Theta_\mathrm{SG}^\top \\ T_\mathrm{IBR}^{-1} \Theta_\mathrm{IBR}^\top \\ T_\mathrm{SG}^{-1} (I - \Lambda) \Theta_\mathrm{SG}^\top
        \end{bmatrix}, \qquad B_{c,d} = \begin{bmatrix}
            0 \\ M^{-1} \\ 0 \\ 0
        \end{bmatrix},
    \end{align}
\end{subequations}
where \(M = \mathrm{diag}(m_i)_{i \in \mathcal{N}}\), \(\Tilde{D} = D - L \, \mathrm{diag}(\mu_i)_{i \in \mathcal{L}} \, F\), with \(D = \mathrm{diag}(d_i)_{i \in \mathcal{N}}\), \(F = -B_{\mathcal{LL}}^{-1}B_{\mathcal{LG}}\) and where \(\mu_i\) denotes the load damping at bus \(i \in \mathcal{L}\), \(T_{IBR} = \mathrm{diag}(\nu_i^\mathrm{IBR})_{i \in \mathcal{N}_{IBR}}\), \(T_{SG} = \mathrm{diag}(\nu_i)_{i \in \mathcal{N}_{SG}}\), \(\Lambda = \mathrm{diag}(\lambda_i)_{i \in \mathcal{N}_{SG}}\), \(\Tilde{K}^\mathrm{pri} = \mathrm{diag}(k_i)_{i \in \mathcal{N}_{SG}}\), and \(S = \Theta_\mathrm{SG} \Lambda \Tilde{K}^\mathrm{pri} \Theta_\mathrm{SG}^\top\), where \(\Theta_\mathrm{SG} \in \mathbbm{R}^{N_I \times N_\mathrm{SG}}\) is an indicator matrix such that \(\Theta_{\mathrm{SG},ij} = 1\) if the \(j\)-th SG is connected to the \(i\)-th bus in the reduced network. In a similar fashion, we define \(\Theta_\mathrm{IBR} \in \mathbbm{R}^{N_I \times N_\mathrm{IBR}}\) such that \(\Theta_{\mathrm{IBR},ij} = 1\) if the \(j\)-th IBR is connected to the \(i\)-th bus, \(M_\mathrm{SG} \in \mathbbm{R}^{N_I \times N_\mathrm{SG}}\) and \(M_\mathrm{IBR} \in \mathbbm{R}^{N_I \times N_\mathrm{IBR}}\) such that \(M_{\mathrm{SG},ij} = m_i\) if the device at the \(i\)-th bus is the \(j\)-th SG, and \(M_{\mathrm{IBR},ij} = m_i\) if the device at the \(i\)-th bus is the \(j\)-th IBR; all other elements of both matrices are zero. We note that our data-driven method does not require knowledge of any of these matrices. \par
The continuous time dynamics may be discretized using a zero-order hold, which is exact when the secondary control inputs are executed in a step-wise fashion; a realistic assumption. Following zero-order hold discretization, the discrete dynamics are represented by:
\begin{equation}
\label{eqn:discrete_dynamics}
    x^{k+1} = Ax^k + B u^{k} + B_d \Delta p^{\mathrm{dist},k},
\end{equation}
where \(k\) denotes the time step. Our goal is to design a controller \(K\) such that \(u^k = Kx^k\).

\subsection{Control-Aware Communication Topology Design Framework}

We model the communication topology defining inter-agent information exchange as a directed graph \(\mathcal{G}_C := \{\mathcal{V}, \mathcal{E}_C\}\), where \((j,i) \in \mathcal{E}_C\) if agent \(j\) sends information to agent \(i\). We assume that each possible link has a cost of communication \(c_{ij} \in \mathbbm{R}_{\geq 0}\). This represents the cost of allocating bandwidth to this communication channel; either by renting bandwidth from a third-party communication provider or utilizing purpose-built communication infrastructure. We aim to optimize \(\mathcal{E}_C\) to balance control performance with the system-wide communication cost, defined by \(\sum_{(j,i) \in \mathcal{E}_C} c_{ij}\). If agent \(i\) does not receive information from agent \(j\), then its control decision cannot be based on information from \(j\); we encode this via an implication constraint \((j,i) \notin \mathcal{E}_C \implies K_{\alpha \beta} = 0 \ \forall (\alpha,\beta) \in \mathcal{I}_{K,ij}\), where \(\mathcal{I}_{K,ij}\) denotes the elements of \(K\) mapping from \(x_j^k\) to \(u^k\). Defining a constraint \(\mathcal{C}_\mathrm{stab}\) that guarantees closed-loop stability when satisfied, and a cost function related to the control cost by \(\Hat{J}_\mathrm{ctrl}(K)\), we arrive at the following problem formulation:
\begin{subequations}
\label{eqn:comms_design_init_formulation}
    \begin{align}
        &\min_{K,\mathcal{E}_C} \Hat{J}_\mathrm{ctrl}(K) + \sum_{(j,i) \in \mathcal{E}_C} c_{ij} \\
        &\mathrm{s.t.} \ \mathcal{C}_\mathrm{stab} \\
        & (j,i) \notin \mathcal{E}_C \implies K_{\alpha \beta} = 0 \ \forall (\alpha,\beta) \in \mathcal{I}_{K,ij} \ \forall i,j \in \mathcal{V}.
    \end{align}
\end{subequations}
Our approach uses data-driven control methods to formulate the stability constraint; this guarantees the existence of a stabilizing controller with a structure consistent with the designed topology, if the controller is designed using the same stability constraint. We use a proxy function for \(\Hat{J}_\mathrm{ctrl}(\cdot)\) to estimate the impact on control performance of including or omitting a particular link to avoid directly optimizing over the closed-loop control cost within the topology design problem, thus improving numerical performance.

\section{Control Design for Distributed Secondary Frequency Regulation with Black-Box Inverters}
\label{section:control_design}


\subsection{Data Collection}


We begin by collecting a dataset which is assembled by a central coordinator. The dataset is collected by injecting a persistently exciting signal into the system; this can be generated by sampling from a uniform distribution. The signal is used to perturb the SG/IBR setpoints and the resulting trajectories of bus angle and frequency as well as SG slow reheat process state and IBR secondary power are measured and communicated to the coordinator. Persistency of excitation conditions to ensure the dataset is sufficiently informative can be found in \cite{Markovsky-Dorfler-2023-Identifiability-Behavioral-Setting} and \cite{Camlibel-Rapisarda-2024-Beyond-Fundamental-Lemma}. We denote dataset inputs and states by \(u^{k,d}\) and \(x^{k,d}\), respectively. For a dataset of length \(N\), we form data matrices \(U = \begin{bmatrix}
    u^{1,d} & \ldots & u^{N-1,d}
\end{bmatrix}\), \(X = \begin{bmatrix}
    x^{1,d} & \ldots & x^{N-1,d}
\end{bmatrix}\) and \(X^+ = \begin{bmatrix}
    x^{2,d} & \ldots & x^{N,d}
\end{bmatrix}\). Assuming that the system has stable open-loop frequency dynamics from the secondary input to the state \(x\), data collection will not cause instability or large oscillations; a small power injection is sufficient as long as the persistency of excitation condition is satisfied. It is also possible to collect multiple short trajectories and concatenate them together appropriately.

\subsection{Structured Controller Design via Robust Data-Driven Methods}


We follow the notation in \cite{Berberich-et-al-2023-Prior-Knowledge-Data-Robust-Design} and define matrices \(Z := \begin{bmatrix}
    X^\top & U^\top
\end{bmatrix}^\top\) and \(W = X^+ - A'X - B'U\), where \(A'\) and \(B'\) represent known parts of the dynamics. Whilst we may know the parameters of the SG re-heater process and dynamics as well as line parameters in the continuous time model, it is intractable to map these directly into the discrete time dynamics without knowing the IBR parameters so we assume the dynamics are totally unknown, thus \(A' = 0\), \(B' = 0\) and \(W = X^+\). This uncertainty may be modelled as \(x^{k+1} = B_w \Psi \begin{bmatrix}
    x^{k^\top} & u^{k^\top}
\end{bmatrix}^\top\) where \(B_w = I\). We separate the power disturbance into a zero-mean stochastic process \(d^k\) and an underlying average disturbance \(\epsilon^k\): \(\Delta p^{\mathrm{dist},k} = d^k + \epsilon^k\). Assuming that we know a norm bound on a disturbance sequence resulting from stochastic process, we formulate a data-driven multiplier to bound the uncertain system dynamics using (18) and (34) in \cite{Berberich-et-al-2023-Prior-Knowledge-Data-Robust-Design} and that \(B_w = I\):
\begin{equation}
\label{eqn:learned_multiplier}
    \Tilde{\mathbf{P}}_d := \left\{ \tau_d \left. \begin{bmatrix}
        -ZZ^\top & Z\Xi^{+^\top} \\
        \Xi^+ Z^\top & \overline{d}I - \Xi \Xi^{+^\top}
    \end{bmatrix} \ \right| \ \tau_d \geq 0 \right\},
\end{equation}
where \(\overline{d} \geq \sum_{i=1}^N \lVert d^k \rVert_2^2\). If there is prior knowledge available in the form of a norm bound on the dynamics such that \(\Psi \Psi^\top \preceq \overline{\psi} I\), this can be incorporated via a prior knowledge multiplier (see (29)-(30) in \cite{Berberich-et-al-2023-Prior-Knowledge-Data-Robust-Design}):
\begin{equation}
\label{eqn:prior_multiplier}
    \Tilde{\mathbf{P}}_\mathrm{pr} := \left\{ \tau_\mathrm{pr} \left. \begin{bmatrix}
        -I & 0 \\ 0 & \overline{\psi} I
    \end{bmatrix} \ \right| \ \tau_\mathrm{pr} \geq 0 \right\}.
\end{equation}
A combined multiplier is formed using the multipliers from prior knowledge and learned from data:
\begin{equation}
\label{eqn:define_com_multiplier}
    \Tilde{\mathbf{P}}_\mathrm{com} := \left\{ \tau_d \Tilde{P}_d + \tau_\mathrm{pr} \Tilde{P}_\mathrm{pr} | \tau_d\Tilde{P}_d\in\Tilde{\mathbf{P}}_d,\tau_\mathrm{pr}\Tilde{P}_\mathrm{pr}\in\Tilde{\mathbf{P}}_\mathrm{pr} \right\},
\end{equation}
where $\tau_d,\tau_{\mathrm{pr}}$ have the same definitions as those in \eqref{eqn:learned_multiplier} and \eqref{eqn:prior_multiplier}.
By combining \cite{Berberich-et-al-2023-Prior-Knowledge-Data-Robust-Design} Thm. 1 with \cite{De-Oliveira-2002-Extended-H2-H-inf-Norm-Characterizations-Controller-Parameterizations} Thm. 5, we formulate a set of linear matrix inequalities (LMIs), the solution of which provides a robust stabilizing controller for all systems consistent with the dataset \(\{X,X^+,U\}\) with a guaranteed closed-loop \(\mathcal{H}_2\) norm bounded by \(\gamma\) for the mapping \(d \mapsto e\), where \(e^k = C_ex^k + D_{eu} u^k\) is a performance signal for some chosen state and input weights \(C_e\), \(D_{eu}\):
\begin{subequations}
\label{eqn:controller_design}
    \begin{align}
    \label{eqn:Gamma_bound}
    \mathrm{trace}(\Gamma) &\leq \gamma^2 \\ 
        \begin{bmatrix}
            \Gamma & C_eG + D_{eu} Y \\
            * & G + G^\top - P
        \end{bmatrix} &\succ 0 \\
        \label{eqn:define_upper_left_block_stability_LMI}
        \Phi = \begin{bmatrix}
            B_d B_d^\top - P & 0 \\ 0 & 0
        \end{bmatrix} + \begin{bmatrix}
            0 & I \\ I & 0
        \end{bmatrix}^\top & \Tilde{P}_\textrm{com} \begin{bmatrix}
            0 & I \\ I & 0
        \end{bmatrix} \\
    \label{eqn:stability_LMI}
        \begin{bmatrix}
		        \Phi & * \\
        \begin{bmatrix}
            0 & G & Y^\top
        \end{bmatrix} & -(G + G^\top - P)
    \end{bmatrix} &\prec 0 \\
    \label{eqn:multiplier_constraint}
    \Tilde{P}_\textrm{com} \in \Tilde{\mathbf{P}}_\mathrm{com},\quad
    P,  \Gamma &\succ 0.
    \end{align}
\end{subequations}
Note that \(\tau_d, \tau_\mathrm{pr}\) enter \eqref{eqn:define_upper_left_block_stability_LMI} through \eqref{eqn:multiplier_constraint} due to our definition of the combined multiplier \eqref{eqn:define_com_multiplier}. The SDP \eqref{eqn:controller_design} is convex in decision variables \(\Gamma\), \(P\), \(G\), \(Y\), \(\tau_d\) and \(\tau_\mathrm{pr}\). The controller is recovered by \(K = YG^{-1}\). We can optimize the \(\mathcal{H}_2\) control performance by minimizing \(\gamma\) directly (and replacing \eqref{eqn:Gamma_bound} with \(\mathrm{trace}(\Gamma) \leq \gamma\)); however, a more reliable numerical implementation uses a semi-stochastic bisection search to find the smallest \(\gamma\) such that \eqref{eqn:controller_design} is feasible.\par

A controller synthesized using \eqref{eqn:controller_design} will be unstructured; however, we wish to impose a specific sparsity pattern according to the communication topology. Specifically, we require \(K \in \mathcal{S}\) with \(\mathcal{S} := \{K \in \mathbbm{R}^{m \times n} \ | \ K_{\alpha \beta} = 0 \ \forall (\alpha,\beta) \in \mathcal{I}_{K,ij} \ \mathrm{if} \ (j,i) \notin \mathcal{E}_C\}\). However, since the mapping from decision variables \(G,Y\) to \(K\) is non-convex (\(K = YG^{-1}\)), there is no exact method to map controller structure constraints onto \(Y\) and \(G\). We utilize sufficient conditions detailed in \cite{Gross-2011-Optimized-Distributed-Control-Network-Topology-Design} to allow for general sparsity patterns at the cost of conservatism. In particular, \(K \in \mathcal{S}\) if the following conditions hold:
\begin{subequations}
\label{eqn:structure_constraints}
    \begin{align}
        Y_{\alpha \beta} &= 0 \ \forall (\alpha, \beta) \in \mathcal{I}_{K,ij} \ \mathrm{if} \ (j,i) \notin \mathcal{E}_C \\
        G_{\alpha \beta} &= 0 \ \forall (\alpha, \beta) \in \mathcal{I}_{G,ij} \ \mathrm{if} \ (j,i) \notin \mathcal{E}_C \\
        G_{\alpha \beta} &= 0 \ \forall (\alpha, \beta) \in \mathcal{I}_{G,zj} \ \mathrm{if} \ (j,i) \notin \mathcal{E}_C \land (z,i) \in \mathcal{E}_C,
    \end{align}
\end{subequations}
where \(\mathcal{I}_{G,ij}\) is the set of all state index pairs mapping from \(x_j^k\) to \(x_i^k\). Note that these are linear equality constraints. The reason for utilizing the extended stability and performance parameterizations in \cite{De-Oliveira-2002-Extended-H2-H-inf-Norm-Characterizations-Controller-Parameterizations} within our control design formulation \eqref{eqn:controller_design} now becomes clear; we avoid imposing structural constraints on the Lyapunov matrix, thereby reducing conservatism \cite{Ferrante-et-al-2020-Design-of-Structured-Stabilizers}. A feasible solution satisfying both \eqref{eqn:controller_design} and \eqref{eqn:structure_constraints} gives a stabilizing controller for the dynamics \eqref{eqn:discrete_dynamics} that has a structure consistent with the communication graph edge set \(\mathcal{E}_C\) and an associated closed-loop \(\mathcal{H}_2\) norm upper-bounded by \(\gamma\).\par

The goal of secondary frequency regulation is to steer bus frequencies back to their nominal values following a disturbance, to regulate tie-line power flows on particular transmission lines back to the desired operating point, and to accomplish this according to the availability of secondary power reserves at each participating device. We encode these objectives within our control design scheme by designing the performance signal \(e^k\) through choices of \(C_e\) and \(D_{eu}\). If we let \(C_e = \begin{bmatrix}
    Q_r^\top & 0
\end{bmatrix}^\top\) and \(D_{eu} = \begin{bmatrix}
    0 & R_r^\top
\end{bmatrix}^\top\), the optimal \(\mathcal{H}_2\) controller is a linear-quadratic regulator with state penalty \(Q = Q_r ^\top Q_r \succ 0\) and \(R = R_r^\top R_r \succ 0\). We choose \(Q = \mathrm{diag}(Q_1,Q_2,0)\) and \(R = \mathrm{diag}(r_i)_{i=1:M}\), where \(r_i\) may be selected according to the available secondary reserve of device \(i\). The penalty \(Q_1\) relates to bus angle deviations (penalizing tie-line power flows), whilst \(Q_2\) relates to bus frequency deviations and the zero block means that we do not penalize the SG slow re-heater or IBR power output states within our scheme.

\subsection{Integral Action}


Without knowing the underlying power disturbance signal \(\epsilon^k\), it is not possible to directly calculate the steady-state control input \(u^{ss}\) required to achieve the desired operating point where \(x = x^{op} = 0\). To circumvent this problem, we use integral action to penalize the cumulative frequency error, leading to the input trajectory converging to the \(u^{ss}\) necessary to ensure \(\Delta \omega = 0\). Since \(\Dot{\Delta \theta} = \omega_0 \Delta \omega\), penalizing \(\int \Delta \omega \ \mathrm{d}t\) is simply equivalent to a penalty on \(\Delta \theta\), therefore we modify \(Q\) such that \(Q = \mathrm{diag}(Q_1+Q_\mathrm{IA},Q_2,0)\), where \(Q_\mathrm{IA}\) is a penalty on the frequency integral. A complete extension to penalize the integral of desired tie-line flows is possible by augmenting the model with a state \(z\) and dynamics \(\Dot{z} = \Delta \theta\); a formulation can be found by adapting results in \cite{Wei-2025-Data-Driven-Tracking-Control-Design-Noisy-Data-DC-MG} Thm. 2, however, we do not reproduce these here for brevity.

\section{Data-Driven Control-Aware Communication Topology Design}
\label{section:topology_design}

The secondary control input of each controllable device \(i\) may be written as \(\Delta p_i^{\mathrm{sec,ref},k} = \Tilde{K}_i \Tilde{x}_i^k\), where \(\Tilde{x}_i^k = \begin{bmatrix}
    x_i^{k^\top} & \begin{bmatrix}
        x_j^{k^\top}
    \end{bmatrix}_{j:(j,i) \in \mathcal{E}_C}
\end{bmatrix}^\top\) and \(\Tilde{K}_i\) represents the suitably concatenated relevant sub-matrices of \(K\). Clearly, the choice of \(\mathcal{E}_C\) will influence the secondary control decisions made and therefore the dynamical performance of the system in closed-loop. We assume that engineering analysis has been conducted to approximate the benefit to control performance of including each possible link; the benefit of including \((j,i)\) in the communication topology is quantified by \(\eta_{ij} \in \mathbbm{R}_{\geq 0}\). The proxy function \(\Hat{J}_\mathrm{ctrl}(K)\) is defined by \(\sum_{(j,i) \in \mathcal{E}_C} \eta_{ij}\).

\subsection{From Topology to Controller Structure}

The existence of an edge in the communication topology can be represented using a Boolean variable \(\delta \in \mathbbm{R}^{M \times M}\); we define each entry \(\delta_{ij}\) by:
\begin{equation}
    \delta_{ij} = \begin{cases}
        1 \ \mathrm{if} \ (j,i) \in \mathcal{E}_C \\
        0 \ \mathrm{otherwise}
    \end{cases}.
\end{equation}
We use \(\delta\) to form a sparsity pattern on the controller \(K\) to enforce the communication topology structure. This may be achieved using an implication constraint such that \(\delta_{ij} = 0 \implies K_{\alpha \beta} = 0 \ \forall (\alpha,\beta) \in \mathcal{I}_{K,ij}\). We now have a decision variable to encode our previous constraint that \((j,i) \notin \mathcal{E}_C \implies K_{\alpha \beta} = 0 \ \forall (\alpha,\beta) \in \mathcal{I}_{K,ij}\). This is equivalent to a sparsity constraint \(K \in \mathcal{S}\) where \(\mathcal{S} := \{K \in \mathbbm{R}^{m \times n} \ | \ K_{\alpha \beta} = 0 \ \forall (\alpha,\beta) \in \mathcal{I}_{K,ij} \ \mathrm{if} \ \delta_{ij} = 0\}\). This constraint can be mapped to the controller parameterization variables using \eqref{eqn:structure_constraints}. However, the implication constraints are non-convex, so we use the Big-M relaxation to convexify them. The relaxed constraints are given by:
\begin{subequations}
\label{eqn:structure_constraints_Big-M}
    \begin{align}
        &-\overline{M} \delta_{ij} \leq Y_{\alpha \beta} \leq \overline{M} \delta_{ij} \  \myquad[1] \forall (\alpha, \beta) \in \mathcal{I}_{K,ij} \ \forall i,j \in \mathcal{V} \\
        &-\overline{M} \delta_{ij} \leq G_{\alpha \beta} \leq \overline{M} \delta_{ij} \  \myquad[1] \forall (\alpha, \beta) \in \mathcal{I}_{G,ij} \ \forall i,j \in \mathcal{V} \\
        &-\overline{M} (\delta_{ij} - \delta_{iz} + 1) \leq G_{\alpha \beta} \leq \overline{M} (\delta_{ij} - \delta_{iz} + 1)  \\
        &  \myquad[11.5] \forall (\alpha, \beta) \in \mathcal{I}_{G,zj} \ \forall i,j,z \in \mathcal{V}, \nonumber
    \end{align}
\end{subequations}
where \(\overline{M}\) is a sufficiently large positive scalar. Generally \(\overline{M}\) should be as small as possible to improve numerical solver performance. However, making \(\overline{M}\) too small will impact solution quality by reducing the size of the feasible set.

\subsection{Optimization Formulation}

The existence of a stabilizing controller can be guaranteed if there is a feasible solution to \eqref{eqn:stability_LMI} for some \(P \succ 0\) and \(\tau_d, \tau_\mathrm{pr} \geq 0\). We therefore choose \eqref{eqn:stability_LMI} to be our stability constraint \(\mathcal{C}_\mathrm{stab}\) in the formulation \eqref{eqn:comms_design_init_formulation}. Our approach is \emph{control-aware} as it guarantees the feasibility of a subsequent structured control design problem assuming that the controller is designed using the same technique as that used in the stability constraint; the control designer is aware of the communication designer's strategy (or conceivably vice versa). The control structure is enforced by the inequality constraints \eqref{eqn:structure_constraints_Big-M}. This leads to an optimization program to design the communication topology from data without directly optimizing over the closed-loop control cost:
\begin{subequations}
\label{eqn:topology_design}
    \begin{align}
        &\min_{\delta,P,G,Y,\tau_d,\tau_\mathrm{pr}} \sum_{i,j=1}^M c_{ij} \delta_{ij} - \eta_{ij} \delta_{ij} \\
        \mathrm{s.t.} \ & \delta \in \{0,1\}^{M \times M} \\
        & P \succ 0, \ \tau_d, \tau_\mathrm{pr} \geq 0 \\
        & \eqref{eqn:stability_LMI}, \ \eqref{eqn:structure_constraints_Big-M}.
    \end{align}
\end{subequations}
This problem is a mixed-integer semidefinite program (MISDP) that is convex in its continuous variables. A controller recovered as \(K = YG^{-1}\) will stabilize the system dynamics \eqref{eqn:discrete_dynamics} whilst having a structure consistent with the communication topology \(\delta\). The objective function seeks to trade off communication cost against control performance; if the communication cost of a certain link is greater than the performance benefit, it will reduce the cost to omit this link, and vice versa if the converse is true.

\section{Case Studies}
\label{section:simulations}

\begin{figure}[thbp]
    \centering
    \includegraphics[width=0.8\linewidth]{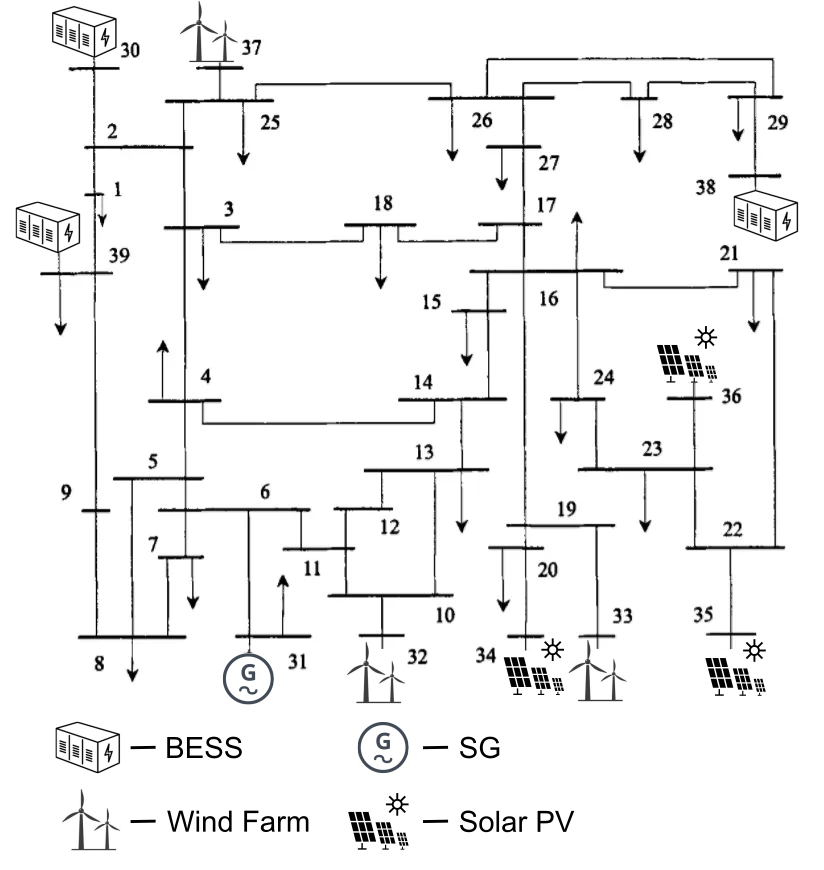}
    \caption{The IEEE 39-bus test system used for our case studies, showing the location of the SG, solar PV units, wind farms and battery energy storage systems (BESS).}
    \label{fig:39_bus_diagram}
\end{figure}

We conduct numerical experiments on a modified IEEE 39-bus test system to validate our communication topology and distributed control design schemes. The physical topology of the test system is shown in Figure \ref{fig:39_bus_diagram}.
An SG is connected to bus 31 whilst GFM/grid-supporting IBRs are connected to buses 
30, 32, 33, 34, 35, 36, 37, 38 and 39. We assume zero load damping for load buses. The system parameters such as inertia, damping, SG parameters and line reactances are randomly generated within realistic bounds. We use a time step of 1 \si{\second}, take \(\overline{\psi} = 300.2\) and set \(\overline{M} = 10^8\). We collect a simulated dataset of 400 samples by injecting a signal sampled from a uniform distribution into the test system. Our control benefit parameters \(\eta = \begin{bmatrix}
    \eta_{ij}
\end{bmatrix}_{i,j \in \mathcal{N}}\) estimate the open-loop coupling between buses by using least-squares regression to identify the state-space model and calculating the normalized Frobenius norm of the relevant submatrix mapping from \(\begin{bmatrix}
    x_j^{k^\top} & \Delta p_j^{\mathrm{sec,ref},k}
\end{bmatrix}^\top\) to \(x_i^{k+1}\): \(\eta_{ij} = \frac{\left\lVert \begin{bmatrix}
    \Hat{A}_{ij} & \Hat{B}_{ij}
\end{bmatrix} \right\rVert_F}{\left\lVert \begin{bmatrix}
    \Hat{A}_{ii} & \Hat{B}_{ii}
\end{bmatrix} \right\rVert_F} \), where \(\Hat{A}\), \(\Hat{B}\) are the estimated state-space matrices. We stress that the stability guarantee of the topology design scheme is independent of the method used to calculate \(\eta\). Naturally, we wish to avoid model identification for determining these parameters in keeping with our direct data-driven approach; however, development of such a method for estimating the control performance benefit of including a given link within the topology is outside the scope of this paper and remains a target for future work. All simulations are conducted in MATLAB, with optimization problems solved using the YALMIP toolbox \cite{Lofberg-2004-YALMIP} and the Mosek solver \cite{mosek} for SDPs. The system operating point is generated using the MATPOWER package \cite{MATPOWER-2011}. \par

To achieve steady-state power sharing between the generators, we calculate the input cost weight \(R\) as \(r_i = \min (\frac{1}{\alpha_i},\overline{r})\), where \(\alpha_i\) is a participation factor calculated from the available power reserve of device \(i\) relative to the total reserve of all devices, and we take \(\overline{r} = 10^3\) to avoid excessively large weights for devices with a small available reserve. Since reserves are only known a day or a few hours ahead, this points towards periodic re-design of the control to account for varying reserve availability. We note that once the controller is designed, online execution is computationally lightweight according to the feedback law \(u^k = Kx^k\). The state cost weight \(Q\) is chosen with \(Q_1 = Q_{1,r}^\top Q_{1,r}\), \(Q_{1,r} = -T_p B_\ell T_\ell F\), \(Q_{IA} = 0.2 I_{N_I}\) and \(Q_2 = 0.8 I_{N_I}\), where \(T_p \in \mathbbm{R}^{n_t \times N_\ell}\) is a matrix selecting which \(n_t\) tie-line power flows to penalize out of the total \(N_\ell\) lines, \(B_\ell \in \mathbbm{R}^{N_\ell \times N_\ell}\) is a diagonal line susceptance matrix, and \(T_\ell \in \mathbbm{R}^{N_\ell \times N_I + N_L}\) is the branch indicator matrix denoting the source and terminal bus indices of each line. This assumes the physical topology and line susceptances are known, although the tie-line power flows can still be penalized if this is not the case as long as power flows are measured and transmitted to the devices managing the lines requiring regulation, with the device states augmented with the tie-line flows.

\subsection{Fully Connected Controller Performance}

\begin{figure}[h]
    \centering
    \includegraphics[width=0.9\linewidth]{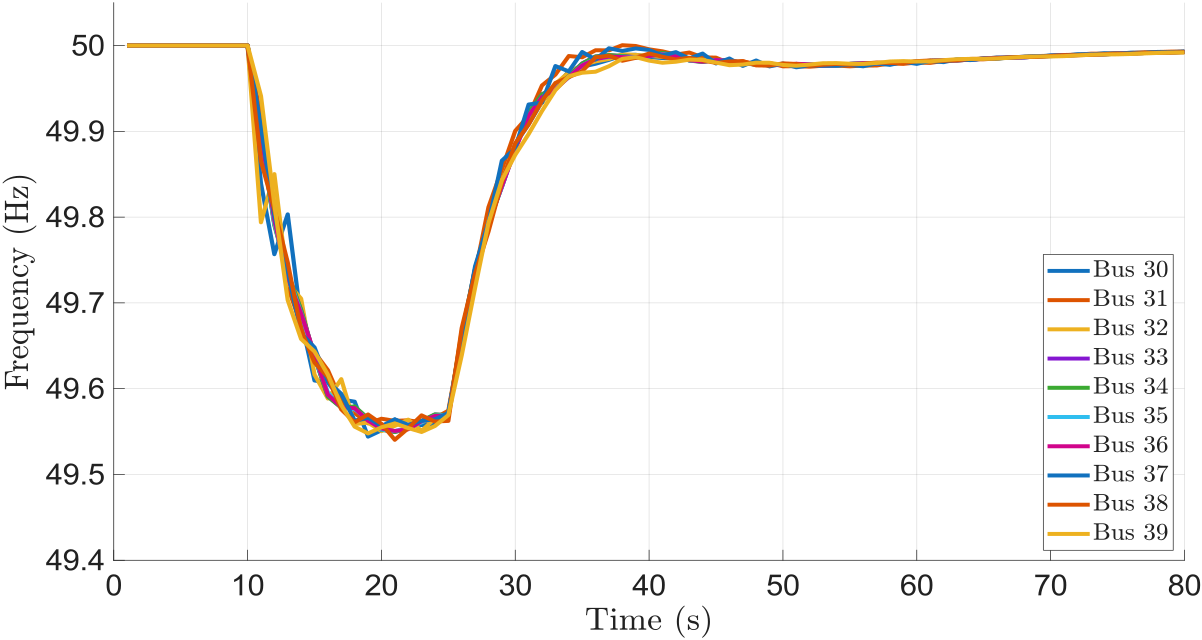}
    \caption{System frequency trajectory evolution is shown against time, with a stable response observed.}
    \label{fig:fully_connected_freq_plot}
\end{figure}

\begin{figure}[h]
    \centering
    \includegraphics[width=0.9\linewidth]{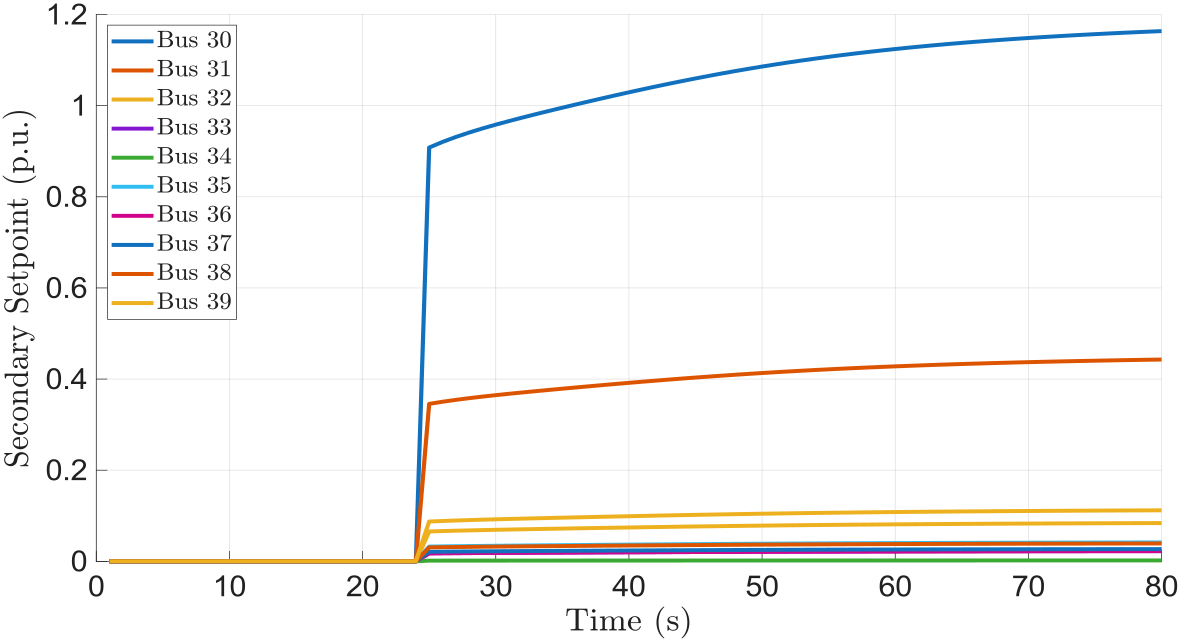}
    \caption{Secondary control input reference plotted against time.}
    \label{fig:fully_connected_ctrl_input_plot}
\end{figure}

We design a dense controller where each control agent communicates with all other agents. This gives the most optimal control performance whilst requiring the maximum level of communication. The secondary controller is designed by solving \eqref{eqn:controller_design} for the smallest feasible \(\gamma\). We simulate a disturbance of \(\epsilon_i^k = -2\) p.u. at bus 31 at \(k = 10\). The controller is activated at \(k = 25\). The resulting closed-loop frequency trajectories are shown in Figure \ref{fig:fully_connected_freq_plot}, whilst the control input trajectories are shown in Figure \ref{fig:fully_connected_ctrl_input_plot}. Zero steady-state error is achieved in the frequency tracking and effective power sharing is observed in the final input values. The frequency recovers to within 10 \% of the nadir value from nominal within 9 \si{\second} of activation of secondary control, with minimal oscillations and no overshoot.

\subsection{Communication Topology Design Study}

\begin{figure}[h]
    \centering
    \includegraphics[width=0.9\linewidth]{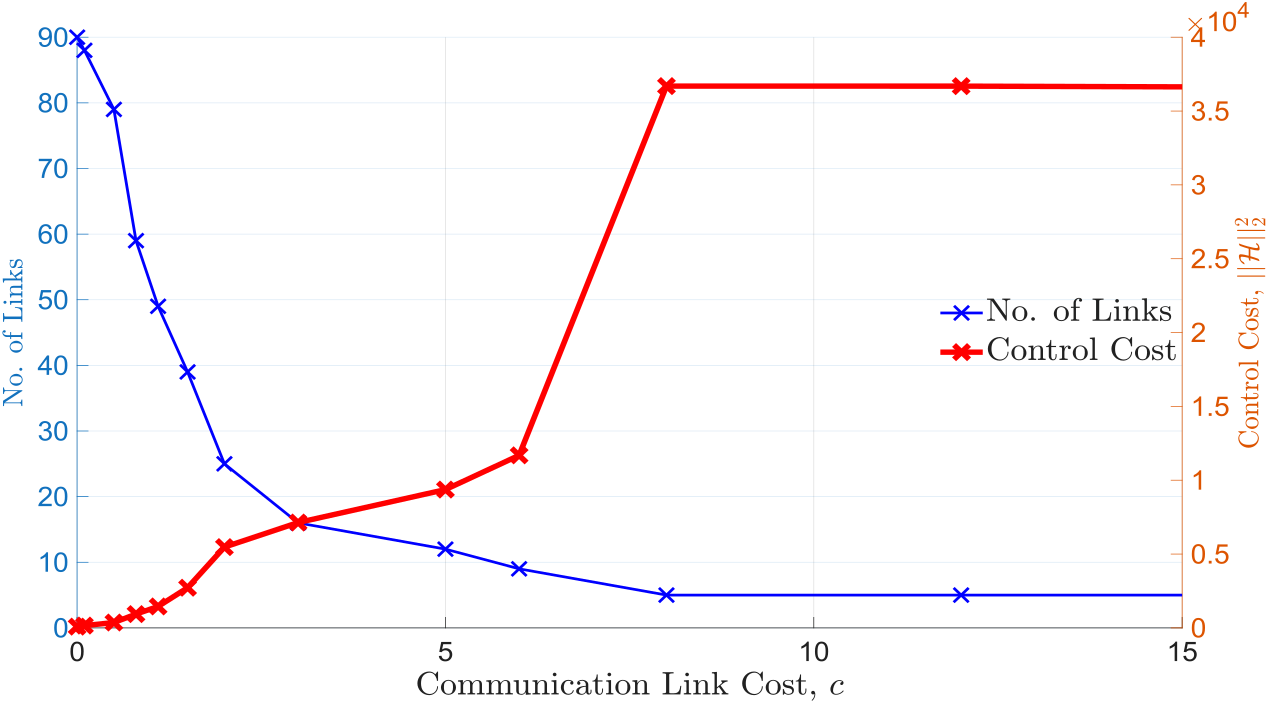}
    \caption{Achieved control cost (green) and number of links in the optimized topology (blue) plotted against the communication link cost, \(c\).}
    \label{fig:ctrl_cost_num_topo_links_against_comms_cost}
\end{figure}

\begin{figure}[h]
    \centering
    \includegraphics[width=0.9\linewidth]{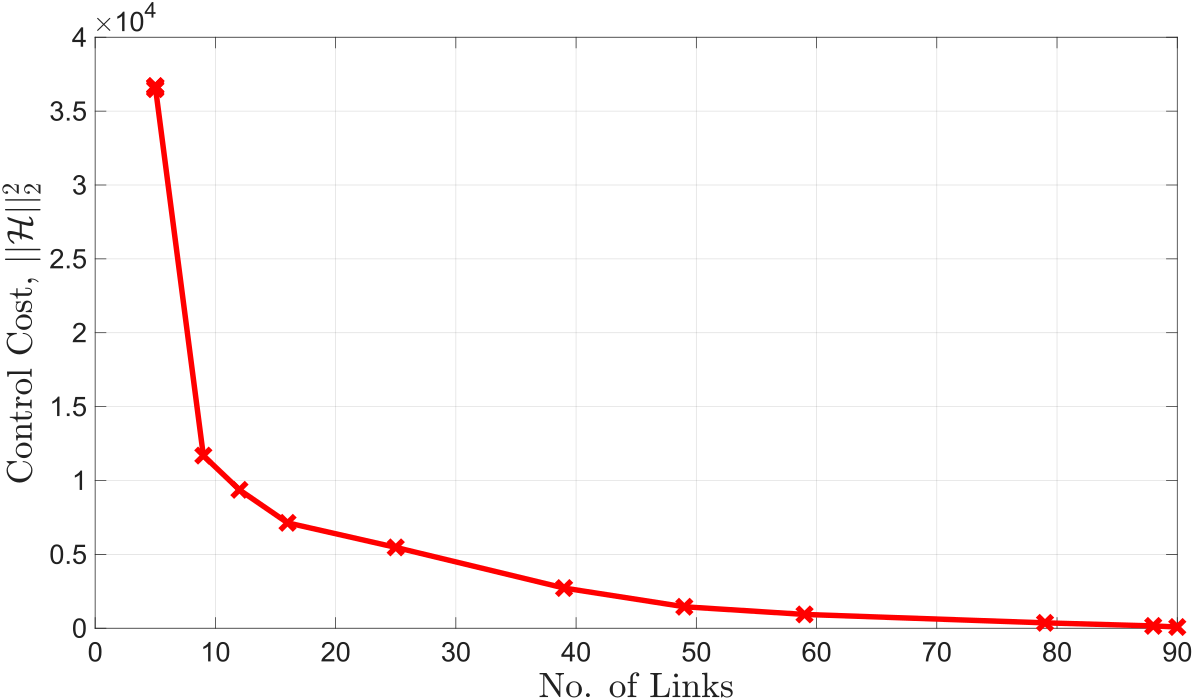}
    \caption{Achieved control cost plotted against the number of links in the communication topology.}
    \label{fig:ctrl_cost_num_topo_links}
\end{figure}

We design a set of communication topologies for a range of communication costs \(c \in \{c_1,\ldots,c_{n_c}\}\), where we set \(c_{ij} = c \ \forall i,j \in \mathcal{N}\). Topology design is executed by solving \eqref{eqn:topology_design}. Subsequently, we jointly solve \eqref{eqn:controller_design} and \eqref{eqn:structure_constraints} for the minimum feasible \(\gamma\) for each topology, where the structural constraints are fixed by the optimized topology, and calculate the squared \(\mathcal{H}_2\) norm of the closed-loop system. It is seen to increase with the communication cost \(c\) for the designed communication topologies and associated controllers in Figure \ref{fig:ctrl_cost_num_topo_links_against_comms_cost}; the figure also shows how the number of links included reduces as \(c\) increases. The relation between the achieved control cost and the number of links in the topology is shown in Figure \ref{fig:ctrl_cost_num_topo_links}. A minimum of 5 links are required to find a stabilizing controller using our formulation; it is likely that the Big-M bound, \(\overline{M}\), restricts the feasible set so that we cannot identify stabilizing controllers with fewer links than this in the topology design step, though they may exist. We note that whilst we use identical cost weights for controller design across the entire set of topologies to enable fair comparison, in reality it is expected that the weights must be re-tuned. This can avoid overly aggressive control policies leading to secondary power reference saturation or poor transient performance.

\subsection{Secondary Power Reference Saturation Response}

Since we take a frequency domain approach to controller design via bounding the \(\mathcal{H}_2\) norm of the system transfer function, we cannot provide theoretical guarantees for satisfying time domain constraints, e.g., input bounds. However, we design the input weights such that the secondary regulation power in steady-state is shared fairly according to the available reserves, so no device will be commanded to exceed its reserve if there is sufficient system-wide reserve and if knowledge of the device reserves is accurate. Whilst the first condition must be met for the secondary regulation problem to be feasible, the second condition cannot be guaranteed. To test the response of our controller in the scenario where reserve knowledge is inaccurate, we conduct a simulation with control weights calculated according to a set of reserve parameters that are different from the actual reserves bounding the device power output.
\begin{figure}[h]
    \centering
    \includegraphics[width=0.9\linewidth]{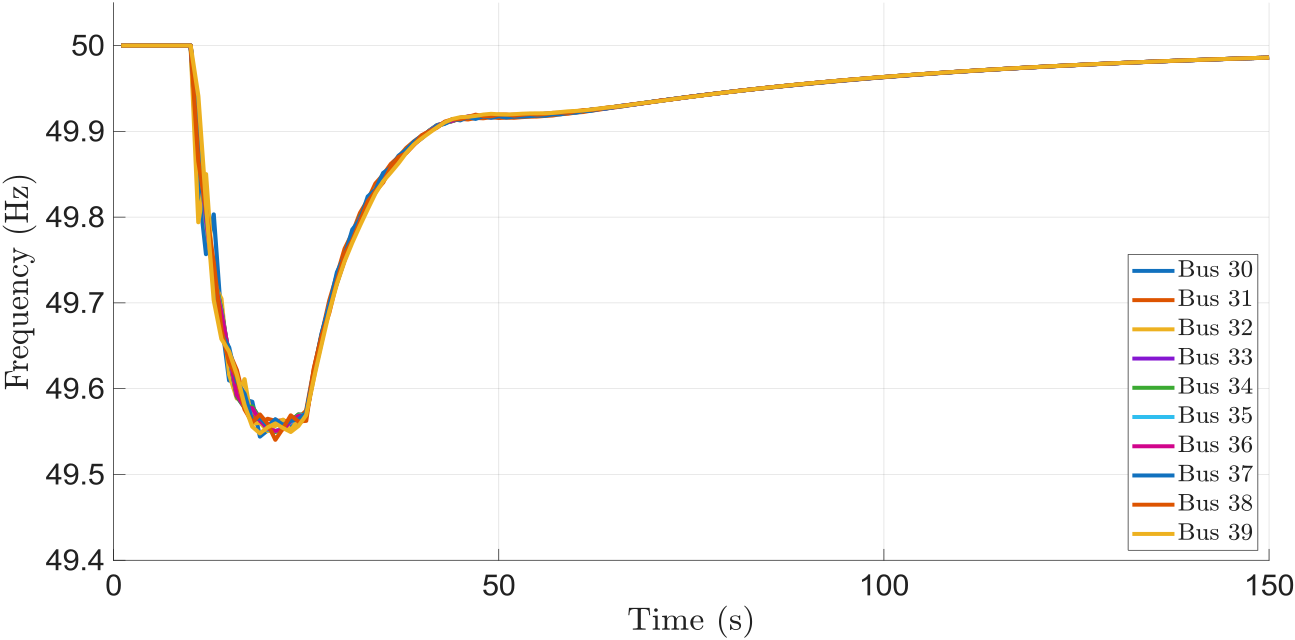}
    \caption{System frequency trajectory evolution for the case with secondary control input saturation.}
    \label{fig:saturated_freq_plot}
\end{figure}
\begin{figure}[h]
    \centering
    \includegraphics[width=0.9\linewidth]{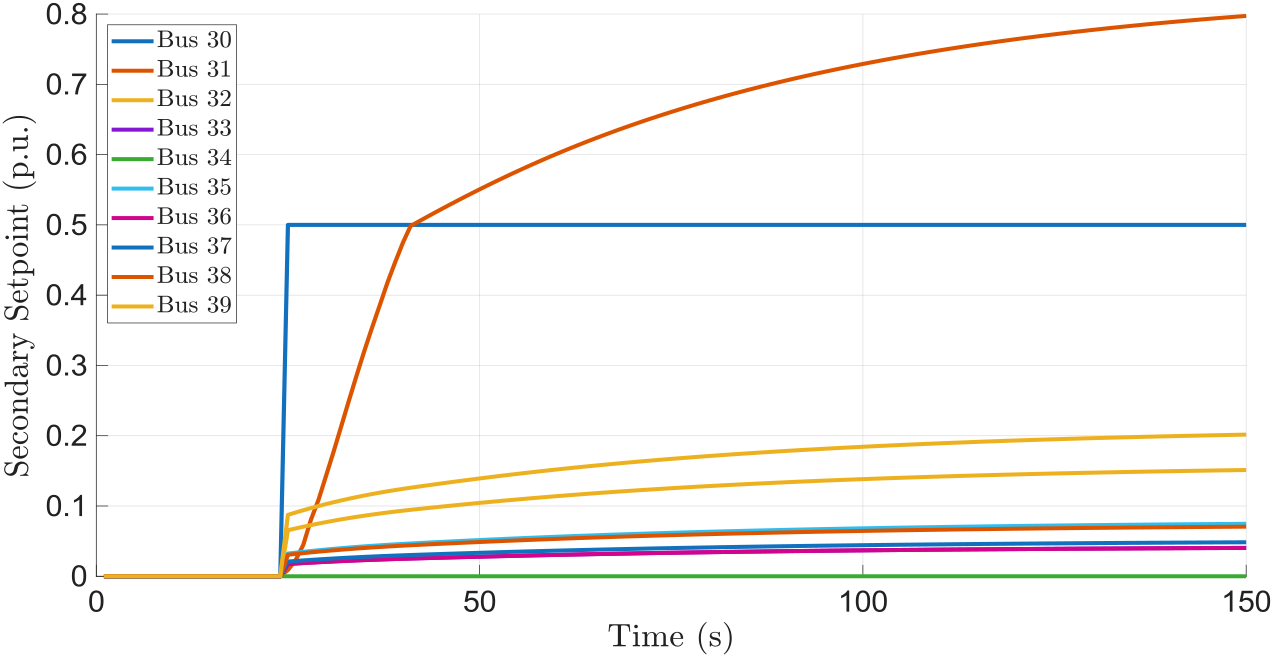}
    \caption{Secondary control input plotted against time, with input saturation at buses 30 and 31.}
    \label{fig:saturated_ctrl_input_plot}
\end{figure}
The controller maintains stability and reaches zero steady-state frequency error, albeit with a slower response compared to the unsaturated case, as shown in Figure \ref{fig:saturated_freq_plot}. The control inputs are shown in Figure \ref{fig:saturated_ctrl_input_plot}; we subtract the varying primary control output from the SG reserve to calculate the maximum secondary regulation capacity of the SG. We see that the BESS unit at bus 30 becomes immediately saturated, whilst the SG secondary input is reduced during the primary response. The remaining IBRs increase their inputs accordingly to regulate the system frequency back to nominal.

\subsection{Experiments with Stochastic Disturbance}

\begin{figure}[h]
    \centering
    \includegraphics[width=0.9\linewidth]{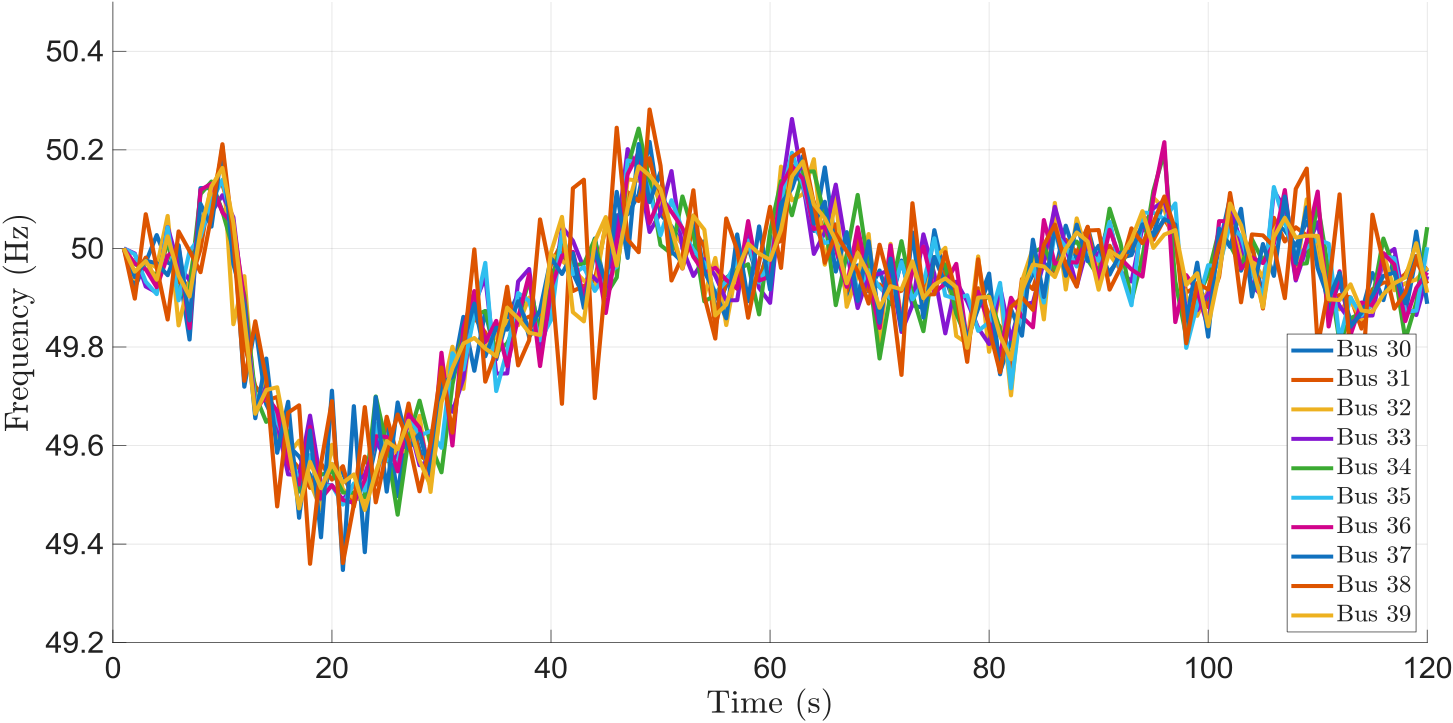}
    \caption{System frequency trajectory evolution with both a step disturbance at 10 \si{\second} and a stochastic process disturbance.}
    \label{fig:stochastic_freq_plot}
\end{figure}

\begin{figure}[h]
    \centering
    \includegraphics[width=0.9\linewidth]{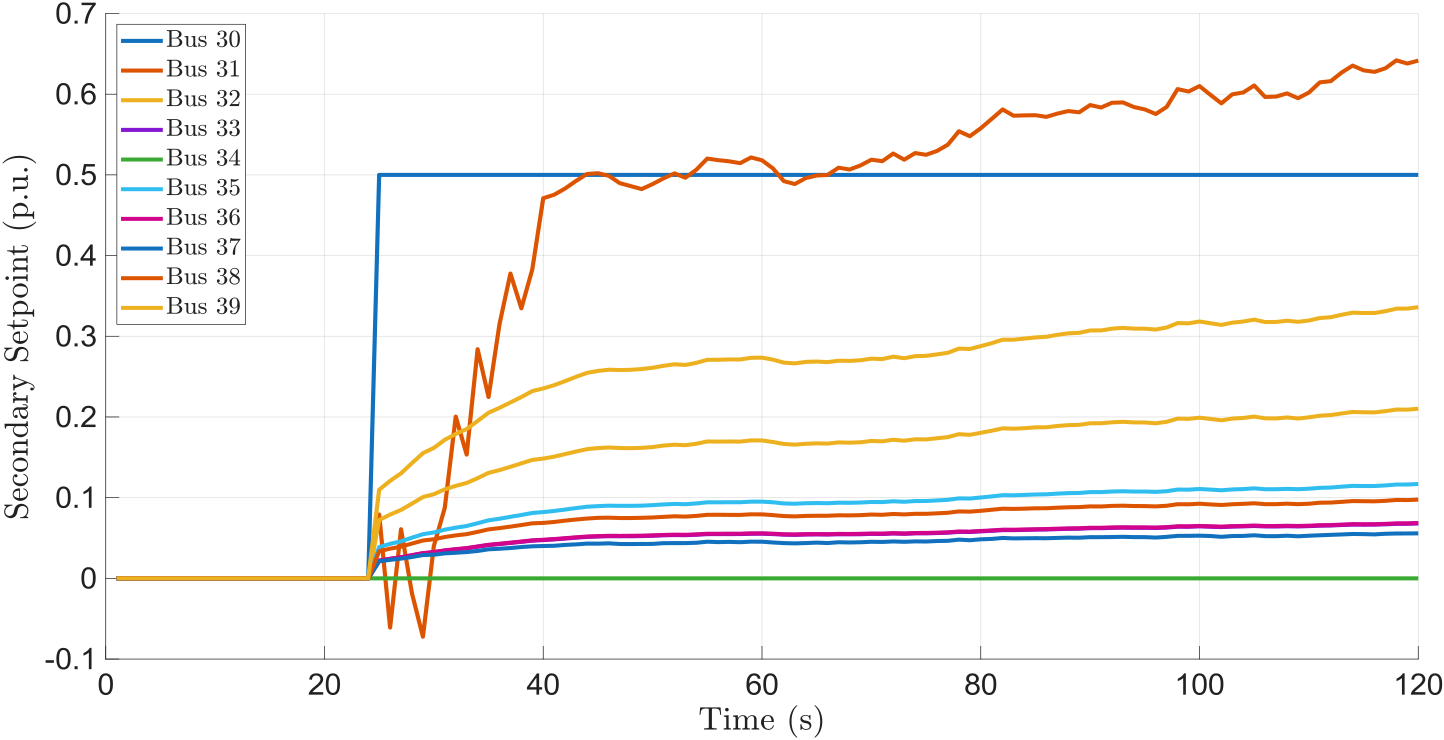}
    \caption{Secondary control input plotted against time for the case with both step and stochastic disturbances.}
    \label{fig:stochastic_ctrl_input_plot}
\end{figure}

Alongside step disturbances, for example, due to a generator tripping, the power grid is also subject to stochastic fluctuations in supply and demand. We carry out simulations where the system is subject to a random power disturbance that is uniformly distributed between \(-0.5\) and \(0.5\) p.u. (\(d^k \sim \mathcal{U}[-0.5,0.5]\)) and is present at each bus with inertia. We also include the step disturbance of \(-2\) p.u. at bus 31, and saturation of the secondary power command sent to each device as outlined above. The controller is successfully able to reject the noise and stabilize the system as shown in Figure \ref{fig:stochastic_freq_plot}, despite the stochastic disturbance being present during data collection. The corresponding control input signals are shown in Figure \ref{fig:stochastic_ctrl_input_plot}.

\section{Conclusion}
\label{section:conclusion}

This paper proposes a two-stage data-driven approach to distributed secondary frequency control. The first stage involves communication topology optimization to design an inter-agent information exchange scheme, whilst in the second stage, a distributed controller is designed, structured according to the topology. The approach provides theoretical guarantees for frequency stability and zero steady-state frequency error. Our sequential method avoids monolithic communication and control design. Numerical experiments demonstrated the effectiveness of the control approach and illustrated the ability of the topology design scheme to trade off between control performance and the cost associated with the communication graph. In future work, we aim to improve the scalability of our approach through distributed optimization, extend the scheme with different control objectives such as mixed \(\mathcal{H}_\infty\)-\(\mathcal{H}_2\) synthesis, and explicitly consider the allocation of device reserves to primary and secondary control.



\bibliographystyle{IEEEtran}

\bibliography{bibliography}

%



\end{document}